# Mechanical Properties of Atomically Thin Tungsten Dichalcogenides: WS$_2$, WSe$_2$ and WTe$_2$

*Alexey Falin,‡[1,2] Matthew Holwill,‡[3,4] Haifeng Lv,‡[5] Wei Gan,[2] Jun Cheng,[1,2] Rui Zhang,[6] Dong Qian,[6] Matthew R. Barnett,[2] Elton J.G. Santos,[7,8] Konstantin S. Novoselov,[3,4,9,10] Tao Tao,[1]\* Xiaojun Wu,[5]\* Lu Hua Li[2]\**

1. Guangdong Provincial Key Laboratory of Functional Soft Condensed Matter, School of Materials and Energy, Guangdong University of Technology, Guangzhou 510006, China.

2. Institute for Frontier Materials, Deakin University, Geelong Waurn Ponds Campus, Waurn Ponds, Victoria 3216, Australia.

3. National Graphene Institute, University of Manchester, Oxford Road, Manchester, M13 9PL, United Kingdom.

4. School of Physics and Astronomy, University of Manchester, Oxford Road, Manchester, M13 9PL, United Kingdom.

5. Hefei National Laboratory for Physical Sciences at the Microscale, School of Chemistry and Material Sciences, CAS Key Laboratory of Materials for Energy Conversion, and CAS Center for Excellence in Nanoscience, University of Science and Technology of China, Hefei, Anhui 230026, China

6. Department of Mechanical Engineering, The University of Texas at Dallas, Richardson, Texas 75080, USA.

7. Institute for Condensed Matter Physics and Complex Systems, School of Physics and Astronomy, The University of Edinburgh, EH9 3FD, United Kingdom.

8. The Higgs Centre for Theoretical Physics, The University of Edinburgh, EH9 3FD, United Kingdom.

9. Department of Material Science & Engineering, National University of Singapore, 117575
   Singapore

10. Centre for Advanced 2D Materials and Graphene Research Centre, National University of
    Singapore, 117546 Singapore

## ABSTRACT

Two-dimensional (2D) tungsten disulfide ($WS_2$), tungsten diselenide ($WSe_2$), and tungsten ditelluride ($WTe_2$) draw increasing attention due to their attractive properties deriving from the heavy tungsten and chalcogenide atoms, but their mechanical properties are still mostly unknown. Here, we determine the intrinsic and air-aged mechanical properties of mono-, bi-, and tri-layer (1-3L) $WS_2$, $WSe_2$, and $WTe_2$ using a complementary suite of experiments and theoretical calculations. High-quality 1L $WS_2$ has the highest Young's modulus (302.4±24.1 GPa) and strength (47.0±8.6 GPa) of the entire family, overpassing those of 1L $WSe_2$ (258.6±38.3 and 38.0±6.0 GPa, respectively) and $WTe_2$ (149.1±9.4 and 6.4±3.3 GPa, respectively). However, the elasticity and strength of $WS_2$ decrease most dramatically with increased thickness among the three materials. We interpret the phenomenon by the different tendencies for interlayer sliding in equilibrium state and under in-plane strain and out-of-plane compression conditions in the indentation process, revealed by finite element method (FEM) and density functional theory (DFT) calculations including van der Waals (vdW) interactions. We also demonstrate that the mechanical properties of the high-quality 1-3L $WS_2$ and $WSe_2$ are largely stable in the air for up to 20 weeks. Intriguingly, the 1-3L $WSe_2$ shows increased modulus and strength values with aging in the air. This is ascribed to oxygen doping, which reinforces the structure. The present study will facilitate the design and use of 2D tungsten dichalcogenides in applications, such as strain engineering and flexible field-effect transistors (FETs).

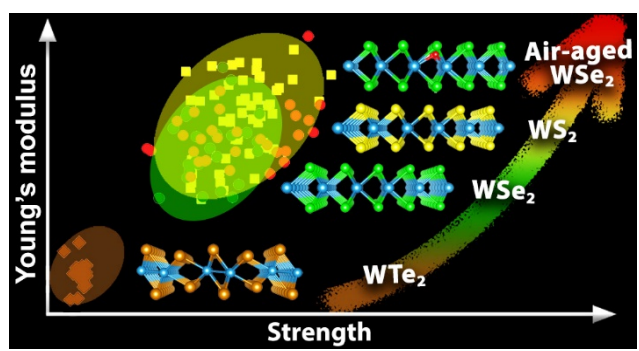

**TOC**

Atomically thin transition metal dichalcogenides (TMDs) is an important group of two-dimensional (2D) materials with bandgaps mostly ranging from 0 to 2.5 eV.[1] Monolayer (1L) molybdenum disulfide ($MoS_2$) sparked intensive attention first due to the discoveries of its transition to direct bandgap[2] and high mobility at room temperature,[3] enabling high-performance flexible light-emitting, photo-detecting, and electronic devices. Then the exploration extended to other 2D TMDs and their wider applications, such as in energy harvesting and storage, chemical sensing, catalysis, and mechanical reinforcement, *etc.*[1]

2D TMDs with heavier transition metals and chalcogens have many advantages. Due to the reduced effective mass, 1L tungsten disulfide ($WS_2$) has been predicted to have the highest mobility and best performance in field-effect transistors (FETs) among 2D TMDs.[4] Thanks to the heavy W atoms, the valence band splitting for 1L tungsten diselenide ($WSe_2$) is about three times larger than that for 1L $MoS_2$, and this renders it more suitable for spintronic and valleytronic applications.[5,6] Quantum spin Hall (QSH) effects have been observed in heavy tungsten ditelluride ($WTe_2$) monolayers because of its most energetically favored semi-metallic 1T' phase,[7-9] unlike the other 2D TMDs in a 2H phase. In addition, huge positive magnetoresistance has been found in $WTe_2$ below 5 K, and more intriguingly, the magnetoresistance does not saturate even under magnetic fields up to 60 T.[10] The reduced intralayer atomic interactions between heavier transition metals and chalcogens make their chemical modification, doping, and functionalization easier to achieve.[11] Tungsten dichalcogenides also have higher electrical conductivities than molybdenum dichalcogenides.[12] The recent advances in the wafer-scale growth based on metal-organic chemical vapor deposition (MOCVD)[13] and gold-assisted millimeter-scale mechanical exfoliation[14] permit high-quality 2D TMDs to be produced, facilitating their use in various applications.

Understanding the mechanical properties of 2D TMDs is essential to harness their full potential for applications, such as in strain engineering and flexible FETs. Extensive studies have been conducted on 2D molybdenum dichalcogenides. The Young's moduli of 1L and 2L $MoS_2$ prepared by mechanical exfoliation and measured by indentation were reported to be 180±60 $Nm^{-1}$ (270±100 GPa) and 260±70 $Nm^{-1}$ (200±60 GPa), respectively; their fracture strengths were 15±3 $Nm^{-1}$ (22±4 GPa) and 28±8 $Nm^{-1}$ (21±6 GPa), respectively.[15] The intrinsic Young's moduli of 5-25L $MoS_2$ were also reported.[16] The elasticity of CVD-grown 1L $MoS_2$ measured by indentation was 171±12 $Nm^{-1}$ (264±18 GPa), and this value is comparable to that of the exfoliated sample.[17] However, the strength of few-layer CVD samples measured by *in situ* tensile testing was only ~1 GPa,[18] one order of magnitude smaller than that of the exfoliated $MoS_2$. This is unlikely to be due to the different test geometry; *in situ* tensile tests provide strength values similar to those obtained using indentation.[19] Rather, it is likely that defect levels are playing a role. The strength of 2D materials is highly sensitive to defects; more so than the Young's modulus.[20] The mechanical properties of molybdenum diselenide ($MoSe_2$) and molybdenum ditelluride ($MoTe_2$) have been experimentally studied as well.[21,22]

In contrast, there have been very few studies on the mechanical properties of atomically thin tungsten dichalcogenides. The Young's modulus of CVD-grown 1L $WS_2$ measured by indentation was $177\pm12$ $Nm^{-1}$ ($272\pm18$ GPa).[17] The Young's moduli of 5L, 6L, 12L, and 14L $WSe_2$ prepared by mechanical exfoliation using indentation, and the obtained values were $596\pm23$ $Nm^{-1}$ ($170\pm7$ GPa), $690\pm25$ $Nm^{-1}$ ($166\pm6$ GPa), $1411\pm61$ $Nm^{-1}$ ($168\pm7$ GPa), and $1615\pm56$ $Nm^{-1}$ ($165\pm6$ GPa), respectively.[23] Nevertheless, no fracture strength was provided. The experimental elasticity and strength values of atomically thin $WS_2$, $WSe_2$, and $WTe_2$ are still mostly unknown, despite a large number of theoretical predictions.[24-28] On the other hand, the long-term stability of 2D TMDs is vital to their applications, especially for the heavier tungsten dichalcogenides which might be more prone to oxidation than the corresponding molybdenum dichalcogenides. The aging of 2D $WS_2$, $WSe_2$, and $WTe_2$ grown by CVD and molecular beam epitaxy (MBE) in ambient conditions could cause catastrophic degradation and lead to quenched luminescence and deteriorated mobility.[29-32] The air-aging effect on the mechanical properties of 2D TMDs, including tungsten dichalcogenides, however, has never been ascertained.

Here, we systematically investigate the intrinsic and air-aged mechanical properties of high-quality 1-3L $WS_2$, $WSe_2$, and $WTe_2$ by indentation. With an increased mass of chalcogenide, the elasticity and fracture strength of 1L tungsten dichalcogenides decreases, where $WTe_2$ is notably weaker than $WS_2$ and $WSe_2$. Nevertheless, the ultimate strain is less affected by chalcogenide mass, as those of 1L $WS_2$ and $WSe_2$ are seen to be quite close. Lighter chalcogenides show more reduced Young's modulus and strength with increased layer thickness. To better understand this phenomenon, we combine finite element method (FEM) with *ab initio* vdW-corrected density functional theory (DFT) calculations to reveal the distinct sliding tendencies in $WS_2$, $WSe_2$, and $WTe_2$ under strain. Although atomically thin $WTe_2$ is not stable in air, the mechanical properties of 1-3L $WS_2$ and $WSe_2$ are not dramatically influenced by exposure to air for up to 20 weeks. The Young's modulus and fracture strength of 2D $WSe_2$ increase after air aging, and the underlying mechanism is theoretically explored.

## RESULTS AND DISCUSSION

**Sample preparation and characterization.** All atomically thin tungsten dichalcogenides were mechanically exfoliated on silicon wafers covered by silicon oxide ($SiO_2$/Si) with pre-etched circular micro-wells of 0.7-1.8 μm in diameter (see Methods). The smaller micro-wells were used for $WTe_2$ due to its much lower exfoliation yield over the larger micro-wells. Optical microscopy was employed to locate 1-3L $WS_2$, $WSe_2$, and $WTe_2$ suitable for mechanical tests, as illustrated in Figure 1a, e, and i. The corresponding atomic force microscopy (AFM) images (and height profiles) are displayed in Figure 1b (c), f (g), and j (k), respectively. The measured thicknesses were in the range of 1.0-1.3 nm for the monolayers.

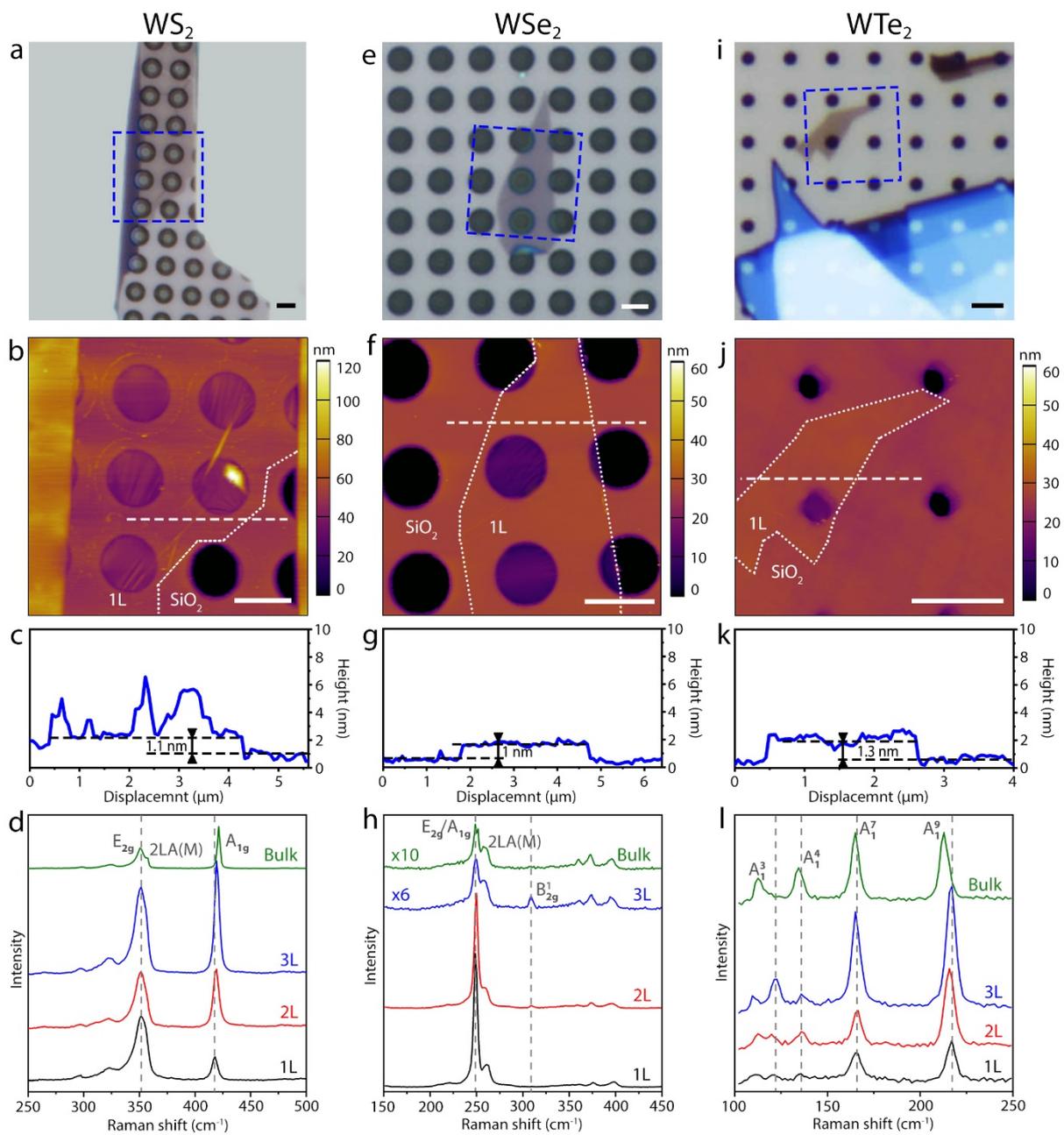

**Figure 1 | Characterizations of the atomically thin tungsten dichalcogenides.** Optical microscopy images of the exfoliated 1L (a) WS₂, (e) WSe₂, and (i) WTe₂ over micro-wells in SiO₂/Si; AFM images of the (b) WS₂, (f) WSe₂, and (j) WTe₂ corresponding to the dashed squares on their optical images; the height traces of (c) WS₂, (g) WSe₂, and (k) WTe₂ corresponding to the dashed lines in their AFM images; Raman spectra of 1-3L and bulk (d) WS₂, (h) WSe₂, and (l) WTe₂ using 514.5 nm excitation wavelength. All scale bars: 2 μm.

The Raman spectra of the $WS_2$, $WSe_2$, and $WTe_2$ for different thicknesses are shown in Figure 1d, h, and l. For $WS_2$, the increased intensity ratio between the $E_{2g}$ peak (in-plane vibrational mode) and $A_{1g}$ peak (out-of-plane vibrational mode) manifested its atomic thickness due to the changed Coulomb interlayer coupling (Figure 1d).[33] Furthermore, the intensity of the 2LA(M) peak increased with thickness reduction and maximized at the monolayer thickness. The $A_{1g}$ peak also experienced redshifts with reduced thickness. For $WSe_2$, the dramatically increased intensity of the $E_{2g}/A_{1g}$ peak verified their atomic thickness (Figure 1h). In addition, 2-3L $WSe_2$ showed $B_{2g}^1$ peaks corresponding to a combination of the rigid interlayer shear and $E_{2g}$ modes, which was not present in 1L or bulk $WSe_2$ (Figure 1h).[34] Distinct from the hexagonal structures of $WS_2$ and $WSe_2$, $WTe_2$ has a distorted orthorhombic phase due to its tilted out-of-plane vibrations caused by metal-metal interaction.[35] $WTe_2$ mainly showed four Raman peaks $A_1^7$ (in-plane vibrational mode） and $A_1^4$, $A_1^3$, $A_1^9$ (out-of-plane vibrational modes) (Figure 1l). The intensities of the $A_1^7$ and $A_1^9$ peaks attenuated with thickness reduction to monolayer, and the $A_1^4$ and $A_1^3$ peaks almost disappeared due to the structural transition from non-centrosymmetric (distorted orthorhombic) to centrosymmetric phase.[36] While the $A_1^7$ peak was almost constant in its position, the $A_1^3$ and $A_1^9$ peaks blueshifted with thickness reduction. To avoid potential damage by laser, all samples measured by Raman were not used in mechanical tests.

**Mechanical properties.** The mechanical behaviors of 1-3L $WS_2$, $WSe_2$, and $WTe_2$ were measured by AFM-based nanoindentation (see Methods).[37,38] The load-displacement relations were obtained from the deflection of the central part of the suspended atomically thin sheets under applied load using cantilevers with diamond tips. The curves were analyzed based on the well-established force ($f$) - displacement ($\delta$) relationship introduced by Lee *et al.*[37] and later extended by Lin *et al.*[39] The relation was employed to determine the starting (zero) coordinates of indentation and the linear and cubic regions of the indentation curve:

$$f = f_0 + k_1(\delta - \delta_0) + k_2(\delta - \delta_0)^3, \qquad (1)$$

where $f_0$ and $\delta_0$ are the zero-point coordinates; $k_1 = \sigma_0^{2D}\pi$, $k_2 = E^{2D}\left(\frac{q^3}{a^2}\right)$, with $\sigma_0^{2D}$ and $E^{2D}$ are the effective pre-tension and effective Young's modulus of a 2D material, respectively; $a$ is the radius of the suspended sheet; $q=1/(1.049–0.15v–0.16v^2)$ is a dimensionless coefficient related to Poisson's ratio ($v$). The Poisson ratios of 1L WS$_2$, WSe$_2$, and WTe$_2$ are 0.22, 0.19 and 0.18, respectively.[24,26] In order to convert the 2D mechanical values to volumetric values, the effective thicknesses of 0.62, 0.65, and 0.71 nm for 1L WS$_2$, WSe$_2$, and WTe$_2$, respectively were adopted.[27] The typical load-displacement curves and corresponding fittings of the three monolayers are compared in Figure 2a.

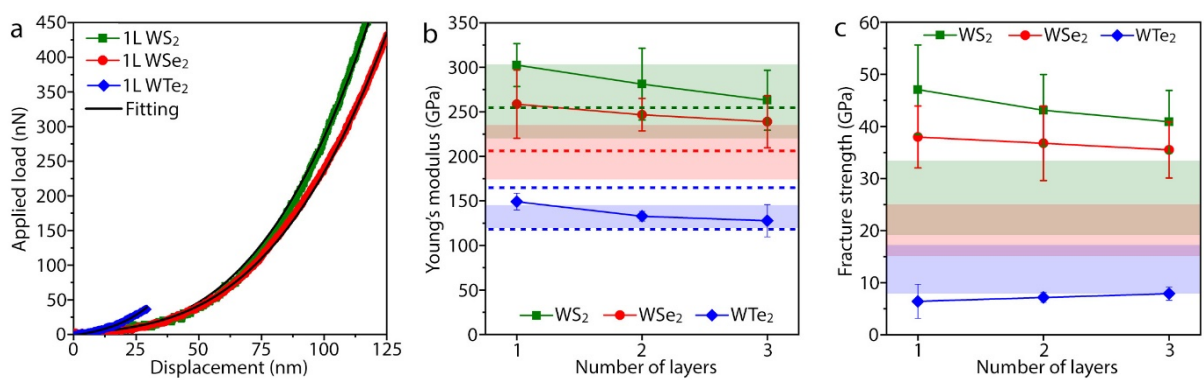

**Figure 2 | Mechanical properties of 1-3L tungsten dichalcogenides.** (a) Typical load-displacement curves of 1L WS$_2$, WSe$_2$, and WTe$_2$ (note WTe$_2$ was suspended over smaller micro-wells), along with the corresponding fittings; (b) the volumetric Young's moduli and (c) fracture strengths of 1-3L WS$_2$, WSe$_2$, and WTe$_2$ (*N*=13, 23, and 16 for 1-3L WS$_2$, respectively; *N*=12, 7, and 6 for 1-3L WSe$_2$, respectively; *N*=3, 5, and 11 for 1-3L WT$_2$, respectively). Colored regions in (b, c) show the ranges of the theoretical values of 1L WS$_2$, WSe$_2$, and WTe$_2$

from literature,[17,24-28] and the dashed lines show the Young's moduli of 1L $WS_2$, $WSe_2$, and $WTe_2$ (in two directions) from our DFT calculations.

The volumetric Young's moduli of the 1-3L tungsten dichalcogenides are shown in Figure 2b, and the values with standard deviations for 1L $WS_2$, $WSe_2$, and $WTe_2$ were 302.4±24.1 GPa (187.5±14.9 $Nm^{-1}$), 258.6±38.3 GPa (168.1±24.9 $Nm^{-1}$), and 149.1±9.4 GPa (105.9±6.6 $Nm^{-1}$), respectively. These values were comparable to the previous theoretical predictions,[17,24-28] illustrated by the colored regions in Figure 2b. We also used DFT to calculate the elastic constants of the three monolayers, and the results were in line with the previous reports (dashed lines in Figure 2b). With the increased number of layers from 1L to 3L, $WS_2$ showed more decreased Young's moduli. This conclusion was confirmed by t-Tests (see Supporting Information, Note 1). The Young's moduli of 3L $WS_2$, $WSe_2$, and $WTe_2$ were 263.1±33.7 GPa (489.4±62.7 $Nm^{-1}$), 238.9±29.4 GPa (465.9±57.3 $Nm^{-1}$), and 127.7±18.1 GPa (272.0±38.0 $Nm^{-1}$), respectively. According to our DFT calculations, 1T' $WTe_2$ should be the most stable phase (Supporting Information, Table S2), consistent with the Raman results. To better show the effect of chalcogens on the mechanical properties, we also calculated and included the elastic constants of 1T-$WS_2$, 1T-$WSe_2$, and 2H-$WTe_2$ in Supporting Information, Table S3.

The fracture strength was calculated by FEM based on the load-displacement curves from the experiment (see Methods). The fracture strength values of the 1L $WS_2$, $WSe_2$, and $WTe_2$ were 47.0±8.6 GPa (29.2±5.3 $Nm^{-1}$), 38.0±6.0 GPa (24.7±3.9 $Nm^{-1}$), and 6.4±3.3 GPa (4.5±2.3 $Nm^{-1}$), respectively (Figure 2c). The strength values of the 1L $WS_2$ and $WSe_2$ were larger than the theoretical predictions, but that of 1L $WTe_2$ was smaller (colored regions in Figure 2c). Note that although no covalent bond formed between the indentation tip and samples even under fracture loads (see Supporting information, Note 4), these strength values should be overestimated due to the frictionless models used in FEM simulations.[40] Similar to the trend observed in Young's modulus, the strength of $WS_2$ dropped more with increased thickness, which was confirmed by t-Tests (see Supporting Information, Note 1). The fracture strength values of 3L $WS_2$ and $WSe_2$ were 40.9±6.0 GPa (76.1±11.2 $Nm^{-1}$) and 35.5±5.4 GPa (69.2±10.5 $Nm^{-1}$), respectively. In contrast, the strength of $WTe_2$ with additional layers showed the opposite trend. It has been reported that the stability of $WTe_2$ decreases with thickness reduction, and 1L $WTe_2$ degrades in the air in just a few minutes.[30] The opposite trend observed in $WTe_2$ was most likely due to its degradation, though its AFM indentations were conducted under constant nitrogen gas flow at 1 L/min. The gas flow did not affect the AFM results according to our comparison tests.

We also calculated the strain distribution using FEM and defined the ultimate strain as the average strain in the material in contact with the indenter at fracture. The strain at the center of the material in contact with the indenter was highly localized and considerably higher than that in the rest of the suspended material that almost followed the linear elastic behavior.[38] The ultimate strains for 1L $WS_2$, $WSe_2$, and $WTe_2$ were 19.8±4.3%, 19.7±4.3%, and 4.4±2.7%, respectively. The thickness increases in $WS_2$ and $WSe_2$ only slightly increased these values, but the ultimate strain for 3L $WTe_2$, *i.e.* 6.8±0.9%, was much larger than that for 1L $WTe_2$, which could be due to more degradation of 1L $WTe_2$ in the air. The ultimate stain values of atomically thin $WS_2$ and $WSe_2$ were comparable to that of 1L graphene but a lot larger than that of 1L BN.[37,38] It suggests the possibility of strain engineering in atomically thin $WS_2$ and $WSe_2$ to large extents without fracture.

**Interlayer interactions**. The elasticity and strength of $WS_2$ showed more decreases with increased thickness than $WSe_2$ and $WTe_2$. The different interlayer interactions in the three materials could be the key to understand the phenomenon. To examine this effect, *ab initio* vdW-corrected DFT calculations on sliding energies under different levels of in-plane strain and out-of-plane compression were performed (see Methods). The strain and compression levels were selected by applying conditions encountered at three radial distances from the indentation center of the suspended 2L sheets under maximum loads.

According to the FEM simulations, the 2L sheets displayed negligible strain at the edge of the micro-wells. That is, far away from the indentation center, the materials were in relaxed equilibrium states, and almost no strain and compression were present. Figure 3a, e, and i show the FEM deduced in-plane nominal strain (solid lines) and out-of-plane hydraulic pressure (dashed lines) distributions close to the indentation center of the three 2L tungsten dichalcogenides at the moment of fracture for a 10.6 nm indenter tip. Close to the contact between the tip and sheet at radial distances of 7.5 nm for $WS_2$, 7.5 nm for $WSe_2$, and 4.2 nm for $WTe_2$ (see the vertical dashed lines in Figure 3a, e, and i), large in-plane strains were applied without out-of-plane compression. Directly under the contact area (*i.e.* radial distances of 0-6 nm for $WS_2$, 0-6 nm for $WSe_2$, and 0-3 nm for $WTe_2$ in Figure 3a, e, and i), the maximum in-plane strain and out-of-plane pressure were reached. The detailed strain and pressure values used in the DFT calculations on the sliding energies of 2L $WS_2$, $WSe_2$, and $WTe_2$ are shown in Figure 3c-d, g-h, and k-l. The sliding energies of $WS_2$ and $WSe_2$ were calculated based on AA' stackings. The sliding energies of $WTe_2$ were calculated based on 1T' structural configuration in the 2H' stacking order. These stacking sequences are the most energetically favored. Two sliding pathways were considered in each material, *i.e.* Path-a and Path-b (Figure 3b, f and j).

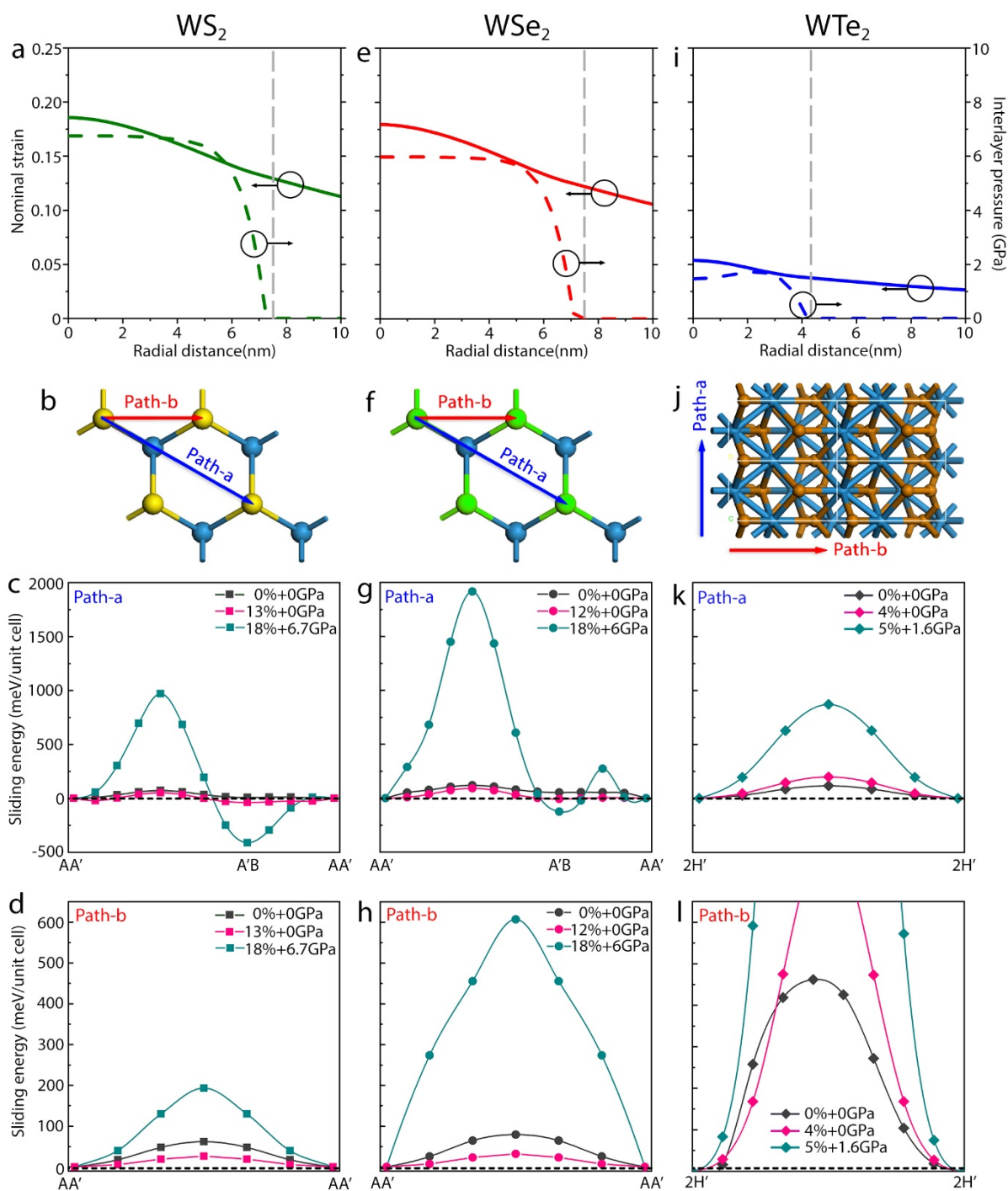

**Figure 3 | Sliding energies of 2L tungsten dichalcogenides under different in-plane strain and out-of-plane compression conditions.** FEM-deduced in-plane strain (solid lines) and out-of-plane pressure (dashed lines) distributions in 2L (a) $WS_2$, (e) $WSe_2$, and (i) $WTe_2$ along a radial distance of 10 nm from the indentation center under fracture loads from an indenter of 10.6 nm in radius; diagrams showing the Path-a and Path-b sliding directions in 2L (b) $WS_2$, (f) $WSe_2$, and (j) $WTe_2$; the sliding energies in Path-a in 2L (c) $WS_2$, (g) $WSe_2$, and (k) $WTe_2$ and in Path-b in 2L (d) $WS_2$, (h) $WSe_2$, and (l) $WTe_2$ under different in-plane strain and out-

of-plane compression conditions, including equilibrium states far away from the indentation center (0% strain + 0 GPa interlayer pressure), large in-plane strain but no out-of-plane compression adjacent to the contact area (*i.e.* 13% strain + 0 GPa pressure for $WS_2$, 12% strain + 0 GPa pressure for $WSe_2$, and 4% strain + 0 GPa pressure for $WTe_2$), and maximum strain and pressure directly under the contact area (*i.e.* 18% strain + 6.7 GPa pressure for $WS_2$, 18% strain + 6.0 GPa pressure for $WSe_2$, and 5% strain + 1.6 GPa pressure for $WTe_2$).

Figure 3c-d, g-h, and k-l reveal how the in-plane strain and out-of-plane compression changed the sliding energies and tendencies in 2L tungsten dichalcogenides. In the relaxed unstrained and unstressed states (*i.e.* 0% strain + 0 GPa compression), $WSe_2$ displayed slightly larger sliding energies (per unit cell) than $WS_2$ in both Path-a and Path-b directions. Note that sliding energy is area-related, and a small change in sliding energy per unit cell can be translated to a significant overall energy barrier in a large area. In the Path-a direction, the maximum sliding energies of $WS_2$ and $WSe_2$, namely 72.0 and 118.6 meV/unit cell, respectively, were present when sliding from AA' (left) to A'B stackings (sliding from left to right in Figure 3c and g). However, the sliding energies from AA' (right) to A'B were as low as ~11.0 and 52.6 meV/unit cell for $WS_2$ and $WSe_2$, respectively (sliding from right to left in Figure 3c and g). In the Path-b direction, the maximum sliding energies were 61.9 and 79.5 meV/unit cell for $WS_2$ and $WSe_2$, respectively (Figure 3d and h). Closer to the indentation center (*i.e.* 7.5 nm radial distance), the 2L $WS_2$ and $WSe_2$ were under 13% and 12% nominal strain, respectively, and no out-of-plane compression was present (*i.e.* 13% strain + 0 GPa compression for $WS_2$ and 12% strain + 0 GPa compression for $WSe_2$). The maximum sliding energies of $WS_2$ and $WSe_2$ decreased in both directions: 51.4 ($WS_2$) and 91.5 ($WSe_2$) meV/unit cell in Path-a (Figure 3c and g) and 26.5 ($WS_2$) and 32.1 ($WSe_2$) meV/unit cell in Path-b (Figure 3d and h), respectively. That is, the maximum sliding energies in $WS_2$ and $WSe_2$ reduced by ~25% and ~58% in Path-a and Path-

b, respectively. Interestingly, the Path-a results indicated that the AA' stacking in $WS_2$ under 13% strain was no longer stable, as the A'B stacking had the minimum sliding energy of −38.6 meV/unit cell (Figure 3c). It means that the 2L $WS_2$ should spontaneously slide from AA' (right) to A'B stacking. In comparison, the AA' stacking was still the most energetically favored position in $WSe_2$, though the sliding energy barrier from AA' (right) to A'B became very small, *i.e.* 3.8 meV/unit cell (sliding from right to left in Figure 3g). Directly under the indenter, the in-plane strain further increased, and additional out-of-plane compression was applied (*i.e.* 18% strain + 6.7 GPa compression for $WS_2$ and 18% strain + 6.0 GPa compression for $WSe_2$). In both Path-a and Path-b sliding directions, the maximum strain and large compression made the maximum sliding energies extremely large, *i.e.* no less than 200 meV/unit cell in $WS_2$ and up to 1917 meV/unit cell in $WSe_2$ (Figure 3c, d, g, and h). Nevertheless, the sliding energy minimum of A'B stacking in $WS_2$ further reduced to −411.5 meV/unit cell (Figure 3d). So spontaneous sliding in $WS_2$ should also occur directly under the indenter. As the sliding leads to strain concentration on the bottom layer and decreases the mechanical properties,[38] the much higher sliding tendencies in $WS_2$ under the three sets of conditions can explain its larger drop in Young's modulus and fracture strength with increased layer number in comparison to $WSe_2$ (Figure 2b and c). The different electronegativity and electronic interactions of S and Se should be the cause for the different sliding tendencies. It should be emphasized that the above-discussed sliding energies are potentially larger than the realistic energy barriers during indentation due to non-uniform stacking changes under indenter.

$WTe_2$ has a different structure to that of $WS_2$ and $WSe_2$, and its sliding energy under strain and compression showed a dissimilar trend as well. In the equilibrium state (*i.e.* 0% strain + 0 GPa compression), the Path-a had a much lower sliding energy than that of Path-b in WTe$_2$: 116.6 *vs.* 461.8 meV/unit cell (Figure 3k and l). With the addition of in-plane strain and out-of-plane compression, the sliding energies in WTe$_2$ showed monotonic increases. At 4.2 nm radial distance from the indentation center, the in-plane strain was 4% but without out-of-plane compression (*i.e.* 4% strain + 0 GPa compression). The sliding energy increased to 199.0 meV/unit cell (Figure 3k). Upon the addition of out-of-plane compression in the contact area (*i.e.* 5% strain + 1.6 GPa compression), the sliding energy jumped to 870.5 meV/unit cell, about 7.5 times that of the equilibrium state (Figure 3k). In Path-b, the sliding energy barrier reached 4000 meV/unit cell under the same strain/compression condition. This suggests that the interlayer interactions in WTe$_2$ were much stronger than those in WS$_2$ and WSe$_2$, possibly due to the strong directional interlayer coupling in addition to the van der Waals interaction, arising from electronic hybridization of the lone electron-pairs and overlap between electronic densities centered at each WTe$_2$ layer.[36] For comparison purposes, we also ran similar DFT calculations on the sliding energies of 2H-WTe$_2$ (Supporting Information, Figure S3).

**Air-aged mechanical properties.** Prior works showed that atomically thin tungsten dichalcogenides could be unstable under ambient conditions.[29-32] We found that the 1-3L WTe$_2$ disappeared after exposure to air for several days; in contrast, the mechanically exfoliated 1-3L WS$_2$ and WSe$_2$ showed almost no morphological change or crack even after 40 weeks under the same condition (Supporting Information, Figure S4). To eliminate the possibility of the WS$_2$ and WSe$_2$ turning to WO$_2$/WO$_3$ that was observed in some studies,[31,41-43] we measured their thickness after air aging by AFM. The thickness of 2D WO$_3$ obtained from oxidation of 1L tungsten dichalcogenides should be 2.2-2.3 nm.[41,43] After 40 weeks in ambient conditions, the thickness of our 1L WS$_2$ and WSe$_2$ increased to ~1.5-1.6 nm, suggesting that these atomically thin $WS_2$ and $WSe_2$ should mostly retain their chemical composition after the long-term air exposure, and the slightly increased thickness should be due to physisorption of airborne organic molecules on their surfaces.[31] We also used Raman spectroscopy to probe the 1-3L $WS_2$ and $WSe_2$ aged in the air for up to 20 weeks (Figure 4). The 1L $WS_2$ showed gradually decreased $E_{2g}$ and $A_{1g}$ peaks with air exposure time (Figure 4a), indicating a certain degree of oxidation.[43,44] In contrast, those of 2-3L $WS_2$ only decreased ~10% after 20 weeks, and the relative intensity of the $A_{1g}$ peak very slightly declined (Figure 4b and c). The $E_{2g}/A_{1g}$ peak intensities of the 1L and 3L $WSe_2$ were stable over the 20 weeks (Figure 4d and f), but that of the 2L dropped with time, and the relative intensity between the $E_{2g}/A_{1g}$ and 2LA(M) peaks also reduced for 2L (Figure 4e). In all aged samples, no new Raman peak at ~800-830 $cm^{-1}$ corresponding to $WO_3$ was detected.[41] These results re-confirmed that most of the mechanically exfoliated atomically thin $WS_2$ and $WSe_2$ were much more stable in the air than the CVD- and MBE-grown samples.[29,31,32] The higher level of oxidation in the 1L $WS_2$ and 2L $WSe_2$ could be due to 1) laser caused oxidation during repeated Raman measurements; 2) potentially more defects that facilitated oxidation.

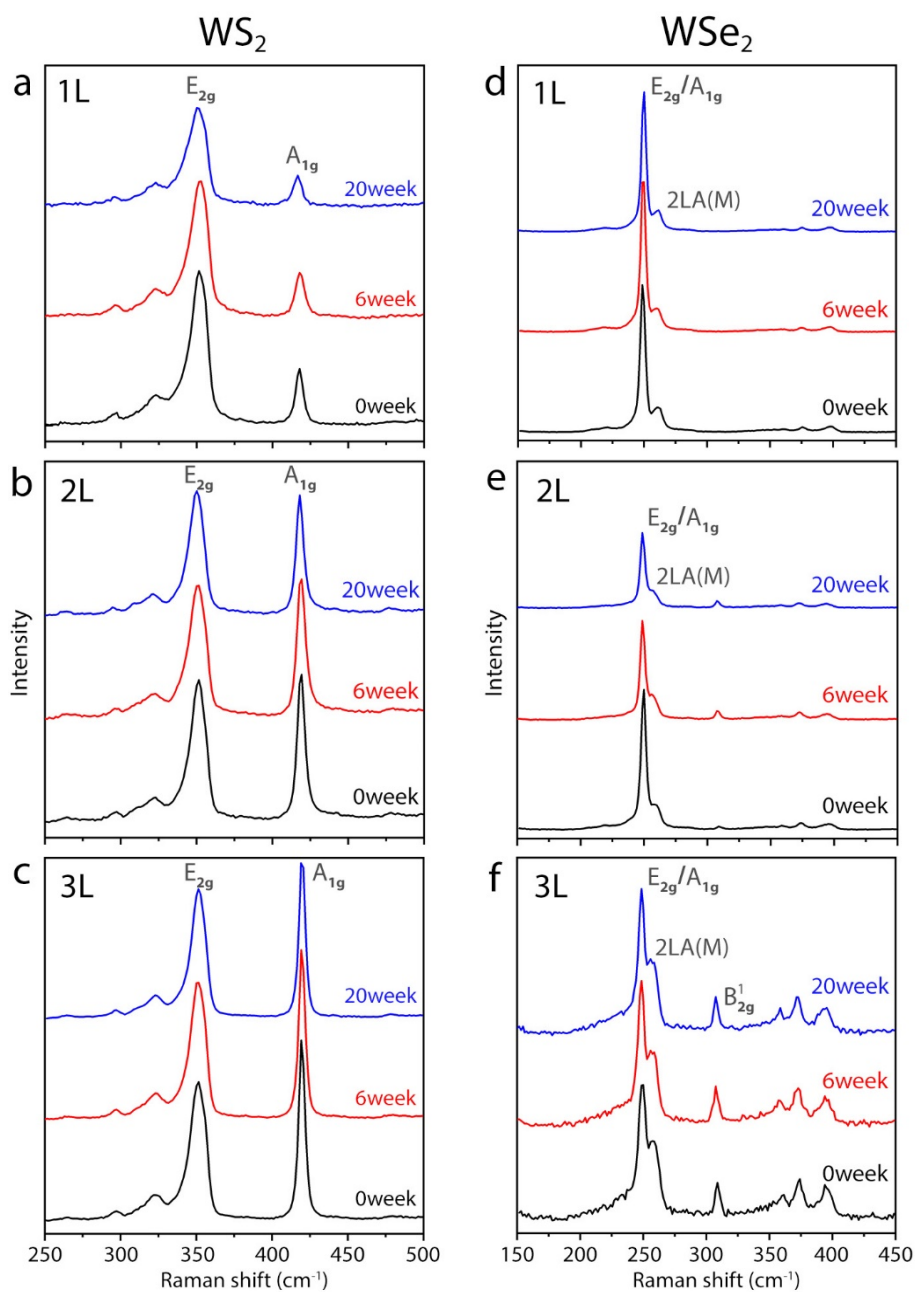

**Figure 4 | Raman characterization of air-aged WS₂ and WSe₂.** Raman spectra of 1-3L (a-c) WS₂ and (d-f) WSe₂ after exposed to air for 6 and 20 weeks.

It is interesting and valuable to test the air-aging effect on the mechanical properties of 2D TMDs, which has never been studied before. Figure 5 summarises the Young's moduli and fracture strengths of the 1-3L WS₂ and WSe₂ after exposure to air for 6 and 20 weeks. Overall, both materials showed no catastrophic mechanical failure, especially that the Young's moduli and fracture strengths were mostly stable within 6 weeks of air exposure, though the standard deviations of strength became much larger. This was consistent with the mostly stable Raman spectra of the aged 1-3L $WS_2$ and $WSe_2$. After 20 weeks, the mean Young's moduli and strengths of 1-3L $WS_2$ decreased, and those of the air-aged 1-3L $WSe_2$ remained stable or even increased in some cases, though statistically there was no significant difference in their means, due to the relatively small number of measurements and the large standard deviations (Supporting Information, Table S4).

Mechanically exfoliated and CVD-grown 2D TMDs tend to have different air-aging effects on their structures. It has been reported that air-aging introduces a large number of vacancies and macroscopic cracks or etching pits to CVD-grown 2D TMDs.[29,32] In contrast, spontaneous oxygen substitution or doping in defect-free basal plane of mechanically exfoliated $MoS_2$ were observed under ambient conditions, and vacancy-type defects were negligible.[45] The difference should be due to the pre-existence of grain boundaries, various vacancies, as well as large amounts of defects in the CVD-grown samples. The vacancies let alone macroscopic cracks or pits have a catastrophic effect on the mechanical properties of air-aged CVD-grown 2D TMDs,[20] but it is not clear how the oxygen substitutional defects in 2D $WS_2$ and $WSe_2$ without the presence of vacancies could affect their mechanical properties. To test the influence of oxygen-doping on the structural and elastic properties of $WS_2$ and $WSe_2$ nanosheets, first-principles calculations were performed based on *ab initio* DFT (Supporting Information, Figure S5). Figure 6 shows the DFT calculated Young's moduli of 1L $WS_2$ and $WSe_2$ at 2.0%, 5.6%, and 12.5% concentrations of oxygen doping (see Supporting Information, Figure S5). Note that the oxygen doping did not change the H-phase of $WS_2$ and $WSe_2$ (Supporting Information, Figure S6 and Table S5). With increased oxygen doping, the elasticity of both materials enhanced, and the averaged lattice parameter *a* decreased, suggesting improved strength as well. Therefore, the mechanical properties of the air-aged atomically thin WS$_2$ and WSe$_2$ should be determined by the combined effects of oxygen doping and vacancy: the former reinforces and the latter deteriorates their mechanical properties. The different trends of the long-term air-aged mechanical properties of WS$_2$ and WSe$_2$ could be due to their different ratios of oxygen doping and vacancy, possibly related to the different density and type of defects in the bulk crystals used for exfoliation.

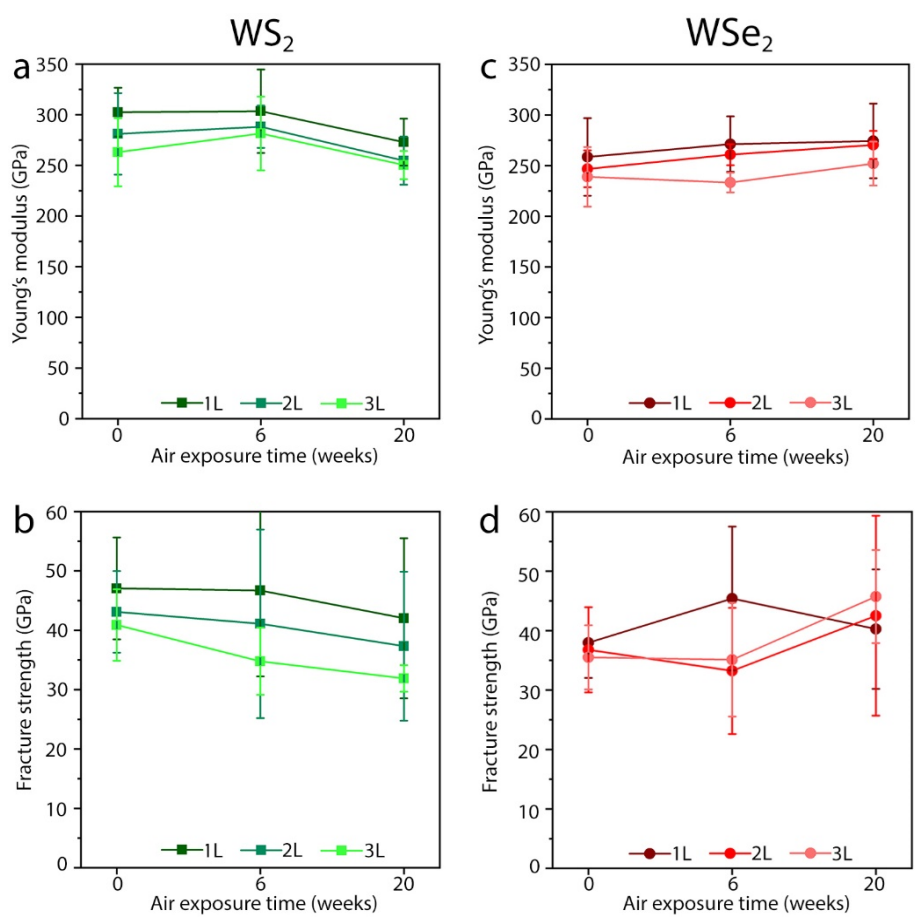

**Figure 5 | Air-aged mechanical properties.** (a, c) Young's moduli and (b, d) fracture strengths of 1-3L WS$_2$ and WSe$_2$ after air exposure for 6 and 20 weeks (*N*=6, 5, and 4 for 1-3L WS$_2$ after 6 weeks, respectively; *N*=4, 4, and 5 for 1-3L WSe$_2$ after 6 weeks, respectively; *N*=4, 9, and 5 for 1-3L WS$_2$ after 20 weeks, respectively; *N*=7, 7, and 5 for 1-3L WSe$_2$ after 20 weeks, respectively).

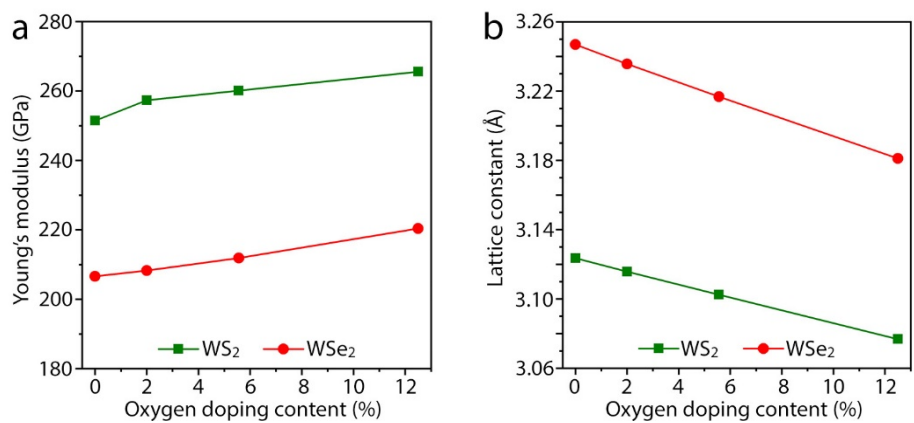

**Figure 6 | Effects of oxygen doping.** DFT calculated (a) Young's moduli and (b) averaged lattice parameters of 1L WS₂ and WSe₂ with different contents of oxygen doping.

## CONCLUSIONS

We used the indentation technique to study the mechanical properties of high-quality atomically thin tungsten dichalcogenides produced by mechanical exfoliation. The Young's moduli of 1L WS₂, WSe₂ and WTe₂ were 302.4±24.1 GPa (187.5±14.9 Nm⁻¹), 258.6±38.3 GPa (168.1±24.9 Nm⁻¹), and 149.1±9.4 GPa (105.9±6.6 Nm⁻¹), respectively; and their fracture strengths were 47.0±8.6 GPa (29.2±5.3 Nm⁻¹), 38.0±6.0 GPa (24.7±3.9 Nm⁻¹), and 6.4±3.3 GPa (4.5±2.3 Nm⁻¹), respectively. The mechanical values of WS₂ reduced most dramatically with increased thickness among the three materials. To understand this phenomenon, we used FEM and DFT calculations to reveal the effects of in-plane strain and out-of-plane compression on the sliding energies in these materials. It was found that WS₂ had the lowest sliding energies no matter in equilibrium state or under strain and compression, and it could spontaneously slide from AA' to A'B stacking during indentation. Interlayer sliding could lead to concentrated stress on the bottom layers, resulting in reduced mechanical properties. We also investigated the long-term air-aging effect on the mechanical properties of 1-3L WS₂ and WSe₂. According to our AFM and Raman characterizations, most of the 1-3L WS₂ and WSe₂ were stable up to 20 weeks after exposure to air. Their mechanical properties could be enhanced due to the oxygen doping by air aging. Our findings imply the possibility of intentional property modification of 2D TMDs *via* aging in a controlled atmosphere.[46]

## Methods

**Materials fabrication and characterization.** The atomically thin tungsten dichalcogenides were mechanically exfoliated from the corresponding bulk crystals by Scotch tape on 90 nm $SiO_2$/Si with prefabricated circular micro-wells of 350, 800, and 900 nm in radius and 1-2 μm in depth in a glovebox filled with argon gas. The Raman spectra were recorded using a 514.5 nm laser (a maximum power of 50 mW) and 100× objective lens (numerical aperture 0.90) on a Renishaw inVia system. For $WS_2$ and $WSe_2$, the laser power was set to 1%, with an absolute energy of ~0.1 mW; for $WTe_2$, the laser power was set to 0.5%, with an absolute power of ~0.05mW to minimize potential damage. These absolute powers were measured by a Newport 1916-C optical power meter after the laser passed through all the optics and hit the samples. All Raman spectra were calibrated using silicon peak at 520.5 $cm^{-1}$. All samples measured by Raman were not used for mechanical tests to prevent the potential effect of laser-induced degradation. The mechanical properties of $WS_2$ and $WSe_2$ were measured using a Cypher AFM in air, and those of $WTe_2$ were measured in the same instrument with a constant flow of nitrogen gas at 1 L/min to reduce degradation. Cantilevers with diamond tips were chosen for indentation to minimize tip deformation-caused errors. The mechanical responses of the cantilevers were calibrated using the combined Sader and thermal noise method. The loading and unloading speed in the indentations were 0.5 μm $s^{-1}$. The loading-unloading curves with large hysteresis were excluded. Transmission electron microscopy (TEM) was employed to measure the tip apex radii, and the obtained tip radii were re-confirmed by indenting high-quality boron nitride[38] and graphene[37] with known mechanical properties. The tip radii were determined to be 3.3 nm (model DEP30: spring constant ($k$) = 22.8 N/m; length ($l$) = 240 μm; width ($w$) = 35 μm; resonance frequency ($f$) = 188.3 kHz), 10.6 nm (D300: $k$ = 27.5 N/m; $l$ = 124 μm; $w$ = 37 μm; $f$ =289.6 kHz), 20.7 nm (D300: $k$ = 57.0 N/m; $l$ = 134 μm; $w$ = 42 μm; $f$ = 336.5 kHz), and 28.5 nm (NC-LC: $k$ = 104.2 N/m; $l$ = 138 μm; $w$ = 40 μm; $f$ = 442.3 kHz). The different tips showed statistically similar values of strength and stiffness. During air aging, the samples were mostly kept in dark at 20 °C in a room with air conditioner without humidity control.

**Finite element analysis.** The analysis was conducted using the commercial nonlinear finite element code ABAQUS (version 6.14). The monocrystal diamond tips of AFM indenters were modeled as rigid spheres. The tested nanosheets were modeled as axisymmetric circular shells with a radius of 350, 800, and 900 nm. The initial thicknesses of nanosheets were assigned to 0.62×N, 0.65×N, and 0.71×N nm for $WS_2$, $WSe_2$, and $WTe_2$, respectively, where N is the number of layers. Two-node linear axisymmetric shell elements (SAX1) were used with mesh densities varying linearly from 0.1 nm (center) to 5.0 nm (outermost). Frictionless contact was assumed in the modeling of the indenter tip - nanosheet interactions. Displacement-controlled loading was applied to the indenter with a linear ramping profile. The tungsten dichalcogenides were assumed to be nonlinear elastic, and the constitutive relation is expressed as:

$$\sigma = E\varepsilon + D\varepsilon^2 \qquad (2)$$

where $D$ is the third-order elastic constant; $\varepsilon$ is an applied strain. Detailed implementations of nonlinear elastic constitutive model in ABAQUS were described in prior work.[38] The Young's modulus and $D$ values of atomically thin tungsten dichalcogenides were obtained from the experimental results. The fracture (curve termination) points in the simulated load-displacement curves were chosen based on the fracture loads from the corresponding experimental curves. Subsequently, the fracture strength and ultimate strain values were obtained as a volume average of the stress and strain values of the elements that were directly underneath the indenter at the corresponding loading steps. The surface-to-surface contact model was chosen to determine the interlayer pressure by modelling the contact pressure between the indenter and nanosheets.

**DFT calculations.** First-principles calculations were performed based on *ab initio* DFT using VASP 5.4.4 code.[47,48] The local density approximation (LDA) was used to calculate the elastic properties of $WX_2$ (X = O, S, Se, and Te) monolayers and PAW-PBE along with optB88-vdW method was used to calculate the sliding energy for bilayer $WX_2$ (X = S, Se, and Te).[49-51] A well-converged plane-wave cutoff of 550 eV was employed. The atomic coordinates were allowed to relax until the forces on the ions were less than $1x10^{-2}$ eV Å$^{-1}$ and the electronic convergence was set to be $1x10^{-6}$ eV. The reduced Brillouin zone was sampled with a $\Gamma$-centered $18\times18\times1$ k-mesh for monolayer/bilayer systems.[52] A 15 Å vacuum space was used in all calculations to avoid any interactions between the supercells in the non-periodic direction. To calculate elastic constants, we chose eleven compression-tensile structures ranging from $-5\%$ to 5% with an interval of 1% compared with the optimized lattice. The doped $WS_2$ and $WSe_2$ monolayers were constructed using a $2\times2$ supercell (12.5%), $3\times3$ supercell (5.6%) and $5\times5$ supercell (2%) with 1 S atom substituted by 1 O atom. The average lattice parameters for oxygen-doped $WS_2$ and $WSe_2$ monolayers were summarized based on the optimized doped systems. To calculate the sliding energy, we chose several intermediate structures along sliding directions (12 and 6 images for $WS_2$ and $WSe_2$ for Path-a and Path-b, respectively; 6 and 10 images for $WTe_2$ for Path-a and Path-b, respectively). Bi-axial strain and hydraulic pressure were used in the calculations.

ASSOCIATED CONTENT

**Supporting Information**. The Supporting Information is available free of charge on the

ACS Publications website at DOI: xx.xxxx/acsnano.xxxxxxx.

> T-Test statistics; DFT calculations on the most stable phase of $WTe_2$, elastic constants and sliding energies of $2H-WTe_2$, interaction between diamond tip and $WS_2$ ($WSe_2$), and air-aged $WS_2$ and $WSe_2$ and their stable phases; optical microscopy images of the aged $WS_2$ and $WSe_2$.

AUTHOR INFORMATION

**Corresponding Author**

* taotao@gdut.edu.cn; xjwu@ustc.edu.cn; luhua.li@deakin.edu.au

**Author Contributions**

L.H.L. conceived and directed this research. A.F. did all mechanical tests and data analysis. M.H. and K.S.N. provided the atomically thin samples. H.L. and X.W. did the DFT calculations. W.G. did Raman. R.Z. and D.Q. carried out FEM calculations. M.R.B., E.J.G.S., J.C. and T.T. provided suggestions and further analysis during discussions. L.H.L. and A.F. co-wrote the manuscript. All authors have given approval to the final version of the manuscript. ‡These authors contributed equally.

ACKNOWLEDGMENT

L.H.L. thanks the financial support from the Australian Research Council (ARC) *via* Discovery Early Career Researcher Award (DE160100796). E.J.G.S. acknowledges computational resources at CIRRUS Tier-2 HPC Service (ec131 Cirrus Project) at EPCC (http://www.cirrus.ac.uk) funded by the University of Edinburgh and EPSRC (EP/P020267/1); ARCHER UK National Supercomputing Service (http://www.archer.ac.uk) *via* Project d429. E.J.G.S. acknowledges the EPSRC Early Career Fellowship (EP/T021578/1) and the University of Edinburgh for funding support. H.F.L and X.J.W. thank the support from the Super Computer Centre of USTCSCC and SCCAS. The authors gratefully acknowledge the support of this work by the grant from the US National Science Foundation (Award CMMI-1727960) and Eugene McDermott Graduate Fellowship at The University of Texas at Dallas.